\documentclass[10pt]{article}

\usepackage{epsfig}

\def\beq{\begin{equation}}  

\def\eeq{\end{equation}}

\newcommand{\eq}{\begin{equation}}

\newcommand{\eqx}{\end{equation}}

\newcommand{\eqn}{\begin{eqnarray}}

\newcommand{\eqnx}{\end{eqnarray}}

\begin{document}
 
\title{Diffractive Higgs boson production at Tevatron and LHC: an experimental
review}
\author{ C. Royon\thanks{%
CEA/DSM/DAPNIA/SPP, F-91191 
Gif-sur-Yvette Cedex,
France, email royon@hep.saclay.cea.fr}}
\maketitle

\begin{abstract}
We discuss the different models of central diffractive production of the Higgs boson at the
Tevatron and the LHC. We also describe how the models can be tested using 
diffractive production data
being taken at the Tevatron. We finally discuss the advantages of using
diffractive events to reconstruct the mass of the Higgs boson especially at the
LHC. 
\end{abstract}

\section{Introduction and motivation}
The discovery of the Higgs boson is one of the main goals of searches at 
the present and next hadronic colliders, the Tevatron
and the LHC. The few Higgs boson candidates found at the end of LEPII \cite{lep}
motivate the search for Higgs boson production for a Higgs mass in
the 115-120 GeV region, specially at the Tevatron. Since this search will be quite
difficult at Tevatron, it is important to find different ways to search
for Higgs bosons. The LHC is clearly the golden accelerator to
look for Higgs bosons. However, the domain at low masses ($M_H < 160$ GeV)
is the most difficult. In that mass domain, the Higgs boson decays mainly into
$b \bar{b}$, $\tau^+ \tau^-$ (for about 10\% of the events) and into $\gamma
\gamma$ (for about 0.1\% of the events). Because of the high background 
in the $b \bar{b}$ or $\tau^+ \tau^-$ channels, only the $\gamma
\gamma$ one seems to be relevant in the standard case and needs quite
high luminosity (about 50 fb$^{-1}$ to get a 5$\sigma$ discovery) and a precise
calibration of the calorimeter. It is thus important to find other ways of
looking for Higgs bosons in the 115-160 GeV mass range, even at the LHC.  

\begin{figure}
\begin{center}
\epsfig{figure=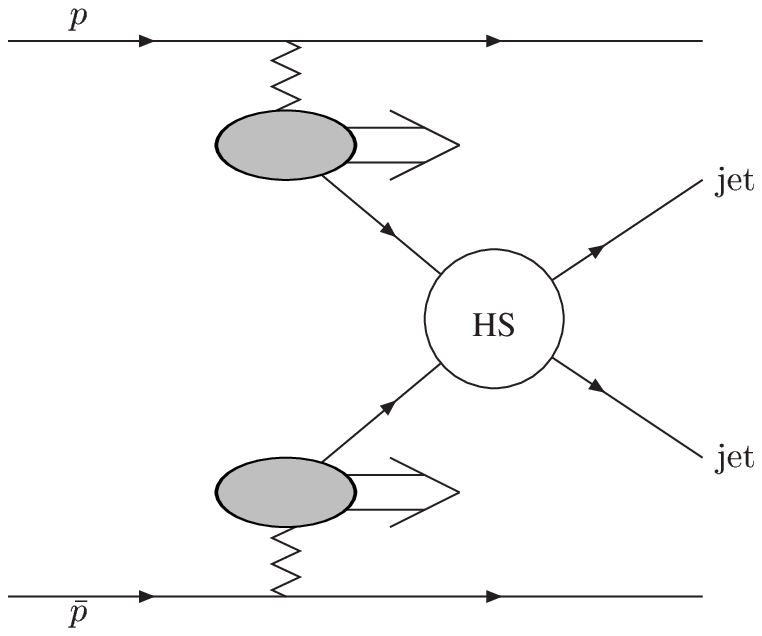,height=1.7in}
\end{center}
\caption{Diffractive jet production via the factorisable inclusive
production mechanism. (Fig taken from Ref. \cite{appleby}).}
\label{diag1}
\end{figure}

\begin{figure}
\begin{center}
\epsfig{figure=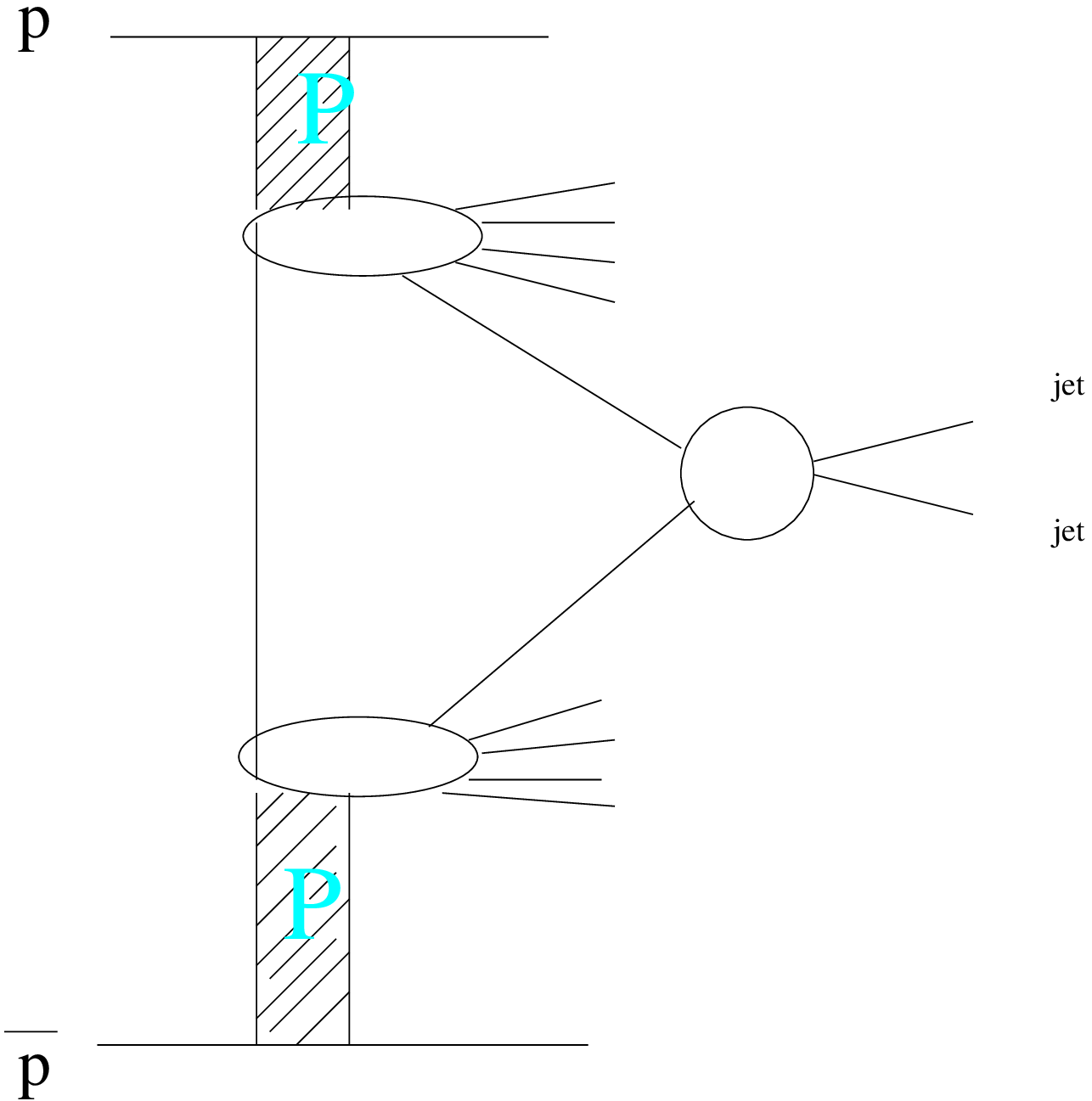,height=1.7in}
\end{center}
\caption{Diffractive jet production via the non factorisable inclusive
production mechanism. (Fig taken from Ref. \cite{pesch}).}
\label{diag2}
\end{figure}

\begin{figure}
\begin{center}
\epsfig{figure=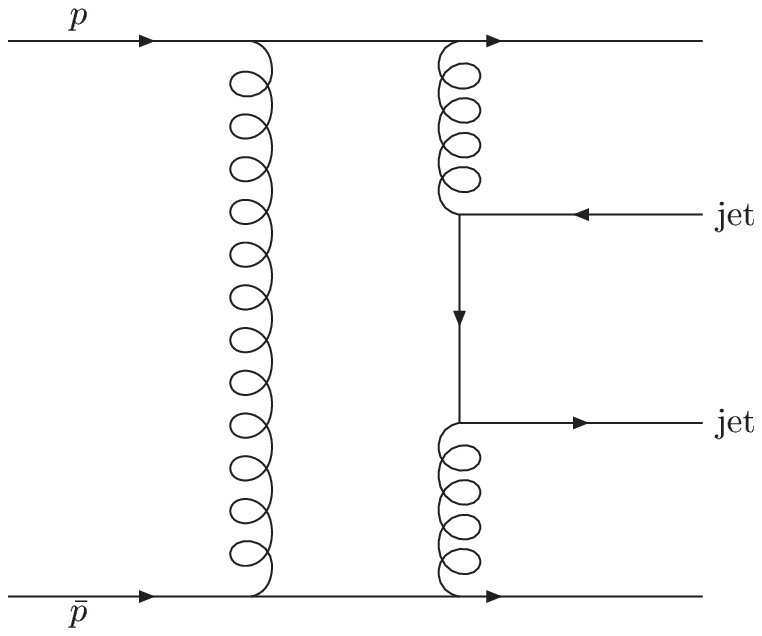,height=1.7in}
\end{center}
\caption{Diffractive jet boson production via the exclusive
production mechanism. (Fig taken from Ref. \cite{appleby}).}
\label{diag3}
\end{figure}

The models of diffractive production of jets, leptons, photons or 
Higgs bosons are given in Figs.
\ref{diag1}, \ref{diag2} and \ref{diag3}. In Fig. \ref{diag1} and \ref{diag2},
we give the schematic production for the inclusive mechanism. As shown in Fig. \ref{diag1},
the diffractive Higgs boson cross section is computed by taking the quark and
gluon densities in the pomeron from the H1 experiment at HERA \cite{h1,cox}.These
densities are then convoluted with the hard subprocess, namely the Higgs boson
production via a top loop. In the second set of models, see Fig. \ref{diag2},
we take the usual hadron-hadron cross section
to produce the hard scattering (Higgs, dijets, $\gamma \gamma$) and
convolute it with the normalised parton densities inside the pomeron taken from the H1
experiment \cite{pesch}. The third set of models \cite{martin} called exclusive 
is completely different since
it does not assume the existence of a colorless object in the proton like
the pomeron, and gives a perturbative calculation of the diffractive process
using the gluon density in the proton (at the lowest order, the exchange of two 
gluons is assumed in order to get a colorless object). We will now discuss all these 
models in more detail, and give their predictions concerning the Higgs boson
production cross section. We will also discuss how to test these models 
using Tevatron data.

\section{Inclusive Higgs production}

\subsection{Factorisable inclusive Higgs production}
The factorisable inclusive Higgs and dijet production cross sections have been
presented in Ref. \cite{appleby, cox}. In this model, the diffractive gluon
density is taken from the H1 measurement \cite{h1} at HERA and is used to compute the
diffractive dijet, diphoton or Higgs boson cross sections, by convoluting this
parton density with the standard hard subprocess (diquark or Higgs boson 
production via a top loop). The cross sections obtained are given in Table
\ref{cox} for the pomeron and reggeon contributions. We will discuss the
assumptions of this model in the following. In this model, factorisation
breaking between HERA and Tevatron is assumed to give a factor 0.1 in normalisation for the Higgs
or dijet production cross sections at the Tevatron. Since this number is not
known at LHC energies, no factor has been applied for the Higgs boson
production cross sections at the LHC. We note that the cross section is 
very low at the Tevatron and quite high at the LHC, provided the survival gap
probability is not too small.

\begin{table}[b]
\begin{center}
\begin{tabular}{|c||c|c|c|c|} \hline
Process &Pomeron& Total &  Total*GSP & GSP\\ 
\hline\hline
$H \rightarrow b \bar{b}$, $m_H=115$ GeV, Tev.  & 0.19 &  0.21 & 0.02 & 0.1 \\
$H \rightarrow WW$, $m_H=160$ GeV, Tev. & 0.003 & 0.003 & 0.0003 & 0.1\\ \hline
$H \rightarrow b \bar{b}$, $m_H=115$ GeV, LHC   & 176 &  276 & 276 & 1. \\
$H \rightarrow WW$, $m_H=160$ GeV, LHC   & 97 & 133 & 133 & 1.
 \\ \hline
\end{tabular}
\caption{Higgs boson production cross section in fb at the Tevatron and the LHC for
Higgs masses between 115 and 160 GeV for the {\it inclusive factorisable} models. The gap
survival (GSP) probability used in this model is also given. At the
LHC, the GSP is taken to be 1.0 since it is not known and all results should
be multiplied by its value (it should be however less than 0.1) 
(see Ref. \cite{cox} for more detail).}
\end{center}
\label{cox}
\end{table}

\subsection{Non factorisable inclusive Higgs production}
The non factorisable inclusive Higgs, dijet, diphoton and dilepton
production cross sections have been presented in Ref. \cite{pesch}. The main
difference with the previous model is that the usual soft hadron-hadron 
cross section is assumed to produce the hard scattering, and this cross section
is then convoluted with the partonic densities in the pomeron taken from the
H1 experiment (see the third item of Ref. \cite{pesch} to get a detailed description
of the theoretical framework of this model). A soft gluon
is present between the two
protons which implies in this model a natural factorisation breaking between
the Tevatron and HERA. The pomeron intercept is thus the soft one in this
model ($\epsilon=0.08$) \cite{bialas} whereas a hard value of the pomeron intercept is
used in the previous model ($\epsilon=0.2$). 

We will also discuss in the following the assumption of taking the quark
and gluon densities from HERA in this model and to apply them to the Tevatron
and the LHC. This model has been interfaced with PYTHIA
\cite{pythia} for hadronisation. The generator
has also been interfaced to a fast simulation of the D\O\ and CDF detectors,
which allowed to scale the prediction to the CDF Run I measurement
\cite{cdf_cross},
namely $\sigma \sim$ 43.6 $\pm$  4.4 (stat.)  $\pm$   21.6 (syst.) nb when 
the antiproton is tagged in the roman pot detectors in the following kinematical
domain in $\xi$ and $t$ of the proton and the antiproton ($0.035 \le \xi_{\bar{P}} \le 0.095$,
$|t| < 1 $ GeV$^2$, $0.01 \le \xi_p \le 0.03$).

\begin{figure}
\begin{center}
\epsfig{figure=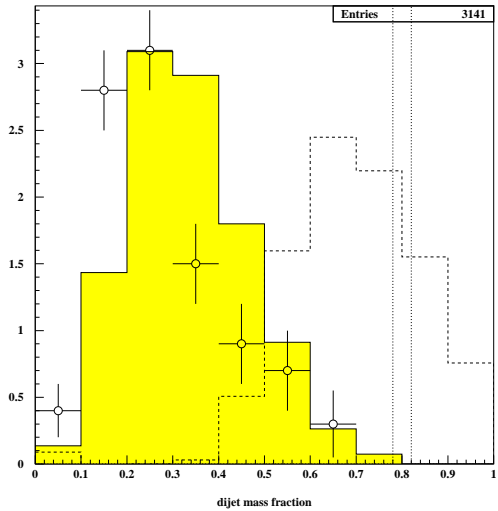,height=3.in}
\end{center}
\caption{Observed dijet mass fraction (CDF Run I), compared to the
non-factorisable pomeron based model 
prediction (full line), using the Pomeron structure functions from H1.
We also give the comparison with exclusive models before (dotted line)
or after (dashed line) simulation of the detector.}
\label{dijetmassfraction}
\end{figure}

\begin{figure}
\begin{center}
\epsfig{figure=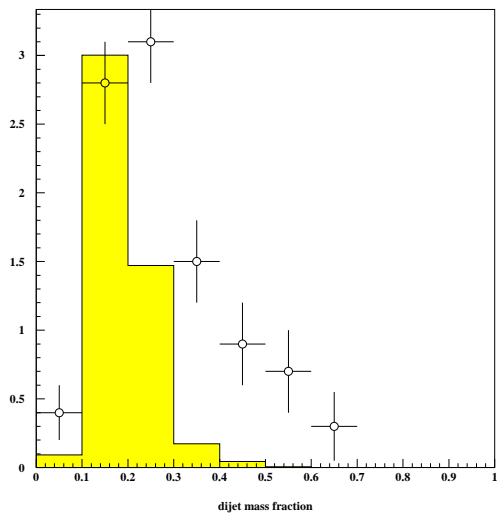,height=3in}
\end{center}
\caption{Dijet mass fraction obtained using the gluon structure function from 
the proton.}
\label{protonsf}
\end{figure}

The prediction of this model has also been compared with the dijet mass fraction
measured by the CDF collaboration \cite{cdf_cross}. The results are shown in
Fig. \ref{dijetmassfraction}. The black points correspond to the CDF
measurements and they are compared with three different curves: the dotted
line gives the prediction at the generator level without any structure function
for the pomeron (bare cross section), the dashed line gives
the same result after simulation of the detector and again no structure assumed
for the pomeron and the full line gives the result of the detector simulation
and the effect of taking the quark and gluon densities in the pomeron from the
H1 experiment. We note that we need both a pomeron made of quarks and gluons,
the parton densities being taken from HERA, and a simulation of the CDF detector
to describe this measurement. Furthermore, if we assume a different structure 
function of the pomeron than the one from H1, namely the same kind of structure 
function as in the proton (we took the GRV \cite{grv} parametrisation to do this
test), we get a bad description of the dijet mass fraction (see Fig.
\ref{protonsf}).

\begin{figure}[p]
\begin{center}
\epsfig{figure=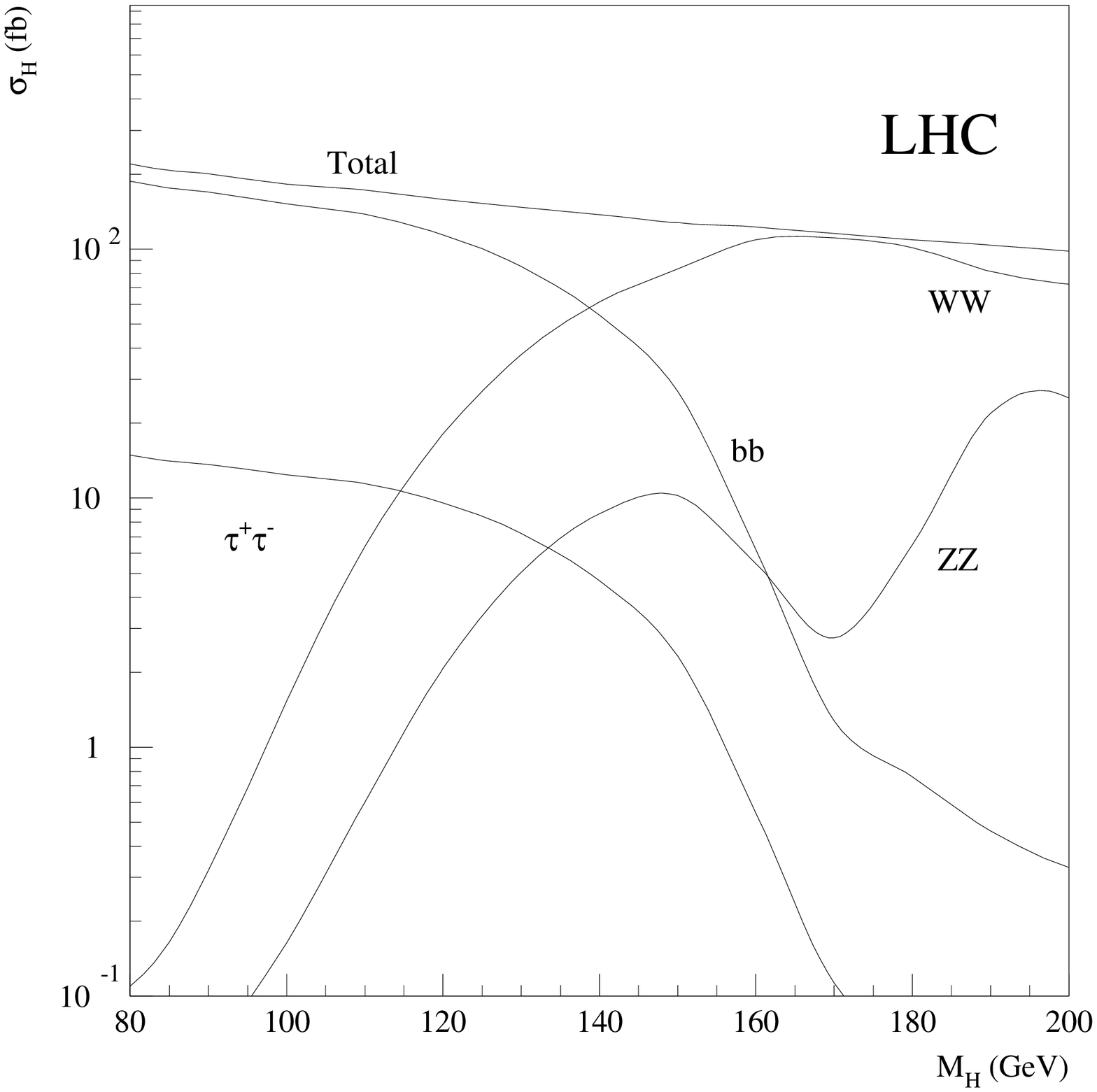,height=3.in}
\end{center}
\caption{ Higgs boson production cross-section at the 
LHC. Various decay channels are plotted as a function of the mass of the 
Higgs boson .}
\label{HiggsXSb}

\end{figure}

The results concerning the Higgs boson cross sections at the Tevatron and the
LHC are given in Table \ref{chris} and in Fig. \ref{HiggsXSb}. We note that the cross sections at
the Tevatron are quite low like in the previous model and 
quite large at the LHC. In this model, there is no need to apply
a survival gap probability since the cross section has been rescaled to the CDF
measurement. It is however important to note that the scaling factor might be
different at the LHC while the same factor between the Tevatron and the LHC
has been assumed.

\begin{table}
\begin{center}
\begin{tabular}{|c||c|c|c|c|} \hline
$M_{Higgs}$&Total&$b \bar{b}$&$\tau \tau$&$W^+ W^-$  \\
\hline\hline
120, Tev. & 1.3 & 0.9  & 0.1  & 0.1 \\
140, Tev. & 0.3 & 0.1  & 0.0  & 0.2 \\ \hline
120, LHC & 15.9 & 11.4 &  1.0 &  1.8 \\
140, LHC & 13.8 &  5.4 &  0.5 &  6.2 \\
160, LHC & 12.3 &   0.6 &  0.1 & 10.9 \\ \hline
\end{tabular}
\caption{Higgs boson production cross section in fb at the Tevatron and the LHC
for different Higgs masses for the inclusive non factorisable models. 
(see Ref. \cite{pesch}
for more detail).}
\end{center}
\label{chris}
\end{table}

\subsection{Factorisation breaking between HERA and Tevatron?}
\begin{figure}
\begin{center}
\vspace{-2.8cm}
\epsfig{figure=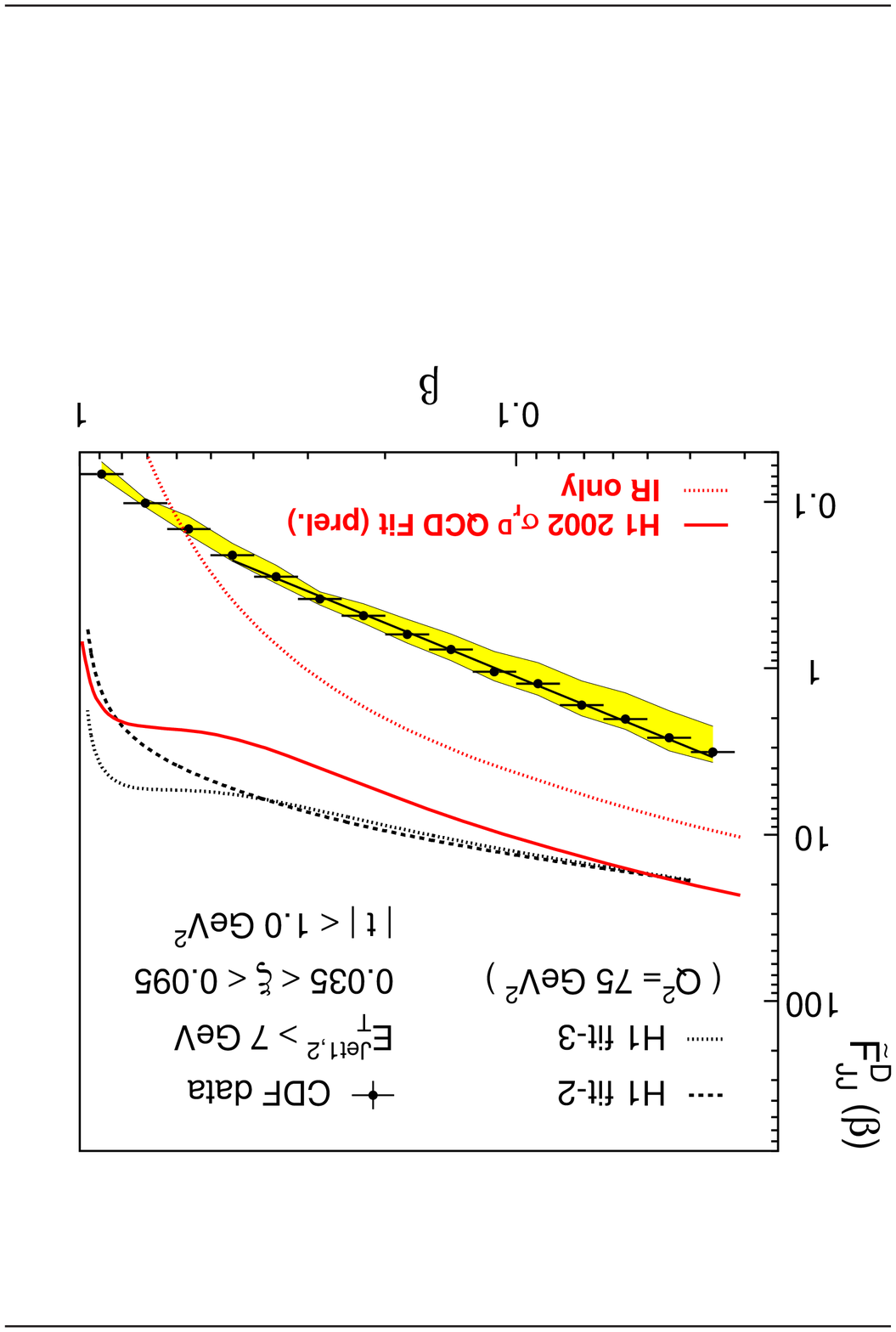,height=7.5in, angle=180}
\end{center}
\caption{Results of the new determination of the gluon density in the pomeron
from the H1 collaboration (full line) compared with the CDF measurement (see Ref. 
\cite{gluonh1} for more detail). (The lower dotted line represents the reggeon
contribution, and the dashed and upper dotted lines the results of a fit using
only 1994 data from the H1 collaboration.)}
\label{fact}
\end{figure}

As we mentionned in the previous sections, one assumption concerning both
models was to take the gluon and quark densities in the pomeron measured at
HERA, in the H1 experiment \cite{h1} to compute the Higgs boson production cross 
section. In Fig. \ref{fact}, we show the recent determination of the gluon
density inside the pomeron made by the H1 collaboration \cite{gluonh1} using
the most recent diffractive structure function data (full red line). The curve
is compared to the CDF gluon density determination \cite{gluonh1} (black
points). We notice that both curves show largely different normalisations but similar
shapes. This justifies a priori to take a constant normalisation factor
difference between HERA and the Tevatron, while keeping the same gluon density.
More precise measurements at the Tevatron being performed using the
CDF roman pot detectors or the new implemented Forward Proton Detector from
D\O\ \cite{fpdd0} will be needed and available soon to test further this
hypothesis.

It is clear however that we need some more studies concerning how the
normalisation factors vary as a function of energy to transfer the results from the
Tevatron to the LHC. One approach to perform this is to study experimentally the
price to pay to get one gap (or one tag in the roman pot detectors) versus
two gaps, since we know already that the percentage of events showing two gaps 
is larger
than the squared percentage of events showing only one gap \cite{dinogap}.

\subsection{SUSY Higgs production cross section enhancement}
\begin{figure}
\begin{center}
\epsfig{figure=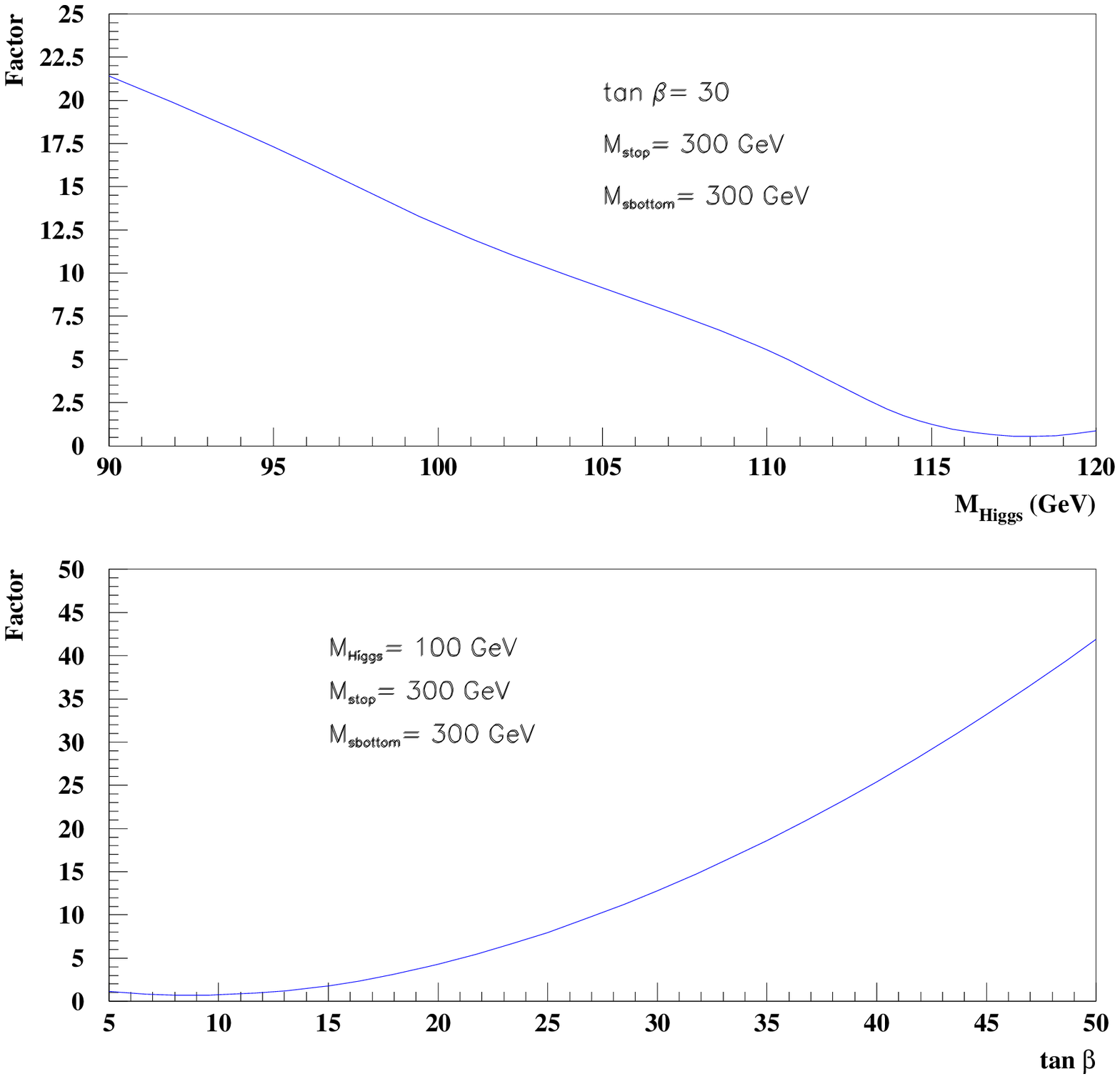,height=4.in}
\end{center}
\caption{Enhancement factor concerning Higgs boson production cross section
in a SUSY scenario compared to the standard model case
as a function of the Higgs mass and the $\tan \beta$ parameter.}
\label{susy}
\end{figure}

Of particular interest are the SUSY based models where mass of the Higgs boson
is expected to be small.
In this section, we will discuss briefly the enhancement in Higgs boson
production cross sections valid for all models (exclusive or inclusive) within
supersymmetric models. At high $\tan \beta$, not only top quark loops have to be 
considered but also bottom quark loops \cite{lavignac}. The results concerning the
Higgs $h$ boson cross section are given in Fig \ref{susy}. The upper plot
gives the enhancement factor as a function of the Higgs mass for $\tan \beta$=30
and the top and sbottom squark masses of 300 GeV in the case where stops are
not degenerate (maximal mixing). We note an enhancement of a factor 10 for
a Higgs mass of 105 GeV for instance. The bottom plot show the distribution
as a function of $\tan \beta$ for a Higgs mass of 100 GeV and we note an
enhancement of the cross section of a factor 40 for $\tan \beta \sim 50$.
In this domain the Higgs boson decays mainly into $b \bar{b}$
(the branching ratio into $\gamma \gamma$ is much smaller than in the case
of the standard model), and it is worth
looking for it in the diffractive channel where $H \rightarrow b \bar{b}$ can be
used. The standard non-diffractive search for Higgs bosons in this mass region
does not benefit from the increase in cross section since the branching ratio
into $\gamma \gamma$ decreases at the same time.

\subsection{Tests of the models at the Tevatron and differences between
the factorisable and non factorisable models}
\begin{figure}[p]
\begin{center}
\epsfig{figure=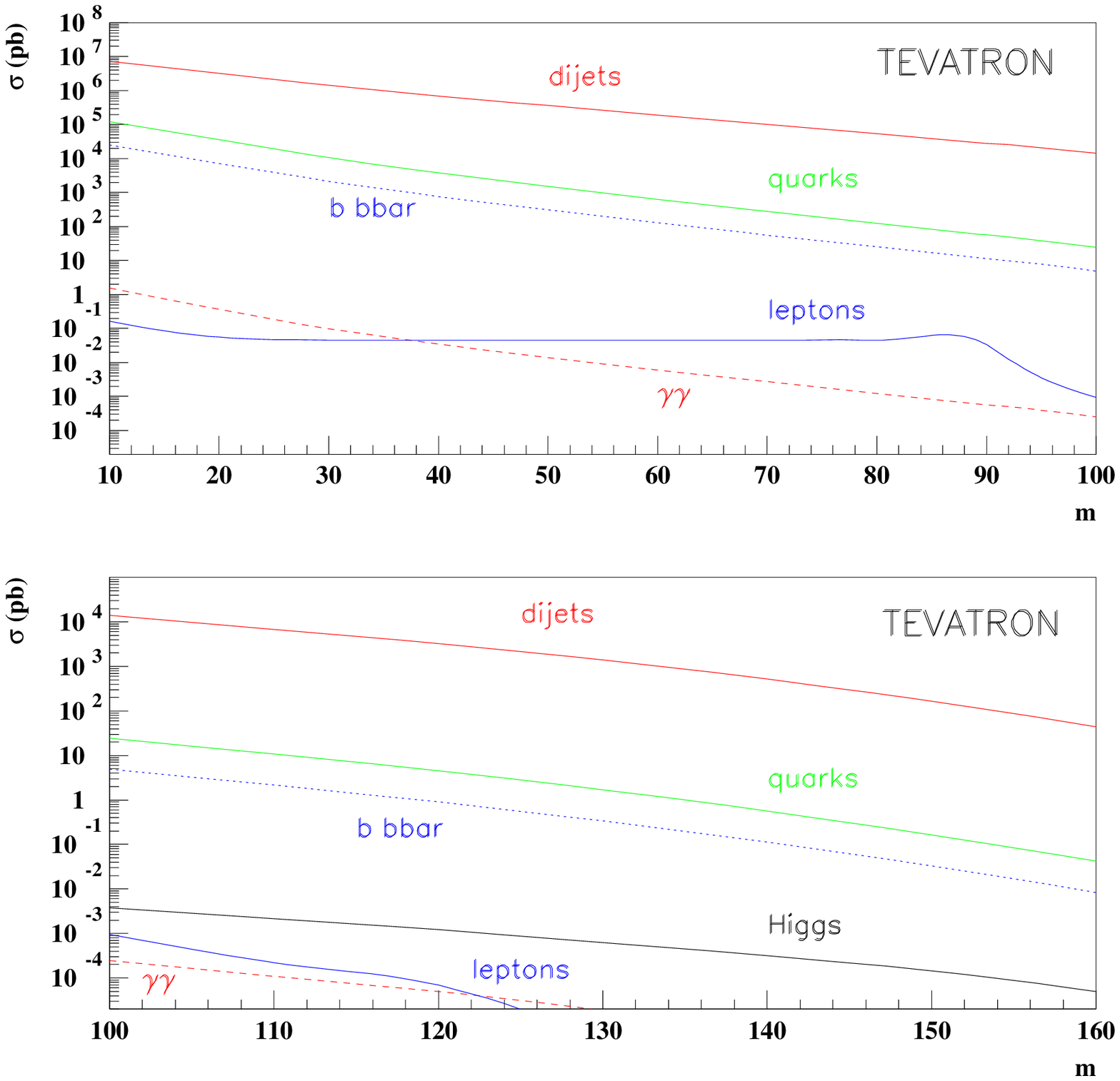,height=6.in}
\end{center}
\caption{Dijet, diquark, $b \bar{b}$, dilepton, and diphoton cross-sections (pb) 
at the Tevatron.
The cross-sections are given for a mass $m$ above the value on the abscissa, for
two different mass ranges. 
For comparison, we also display the Higgs boson cross-section
for $M_{Higgs}=m$.} 
\label{crosster}
\end{figure}

\begin{figure}[p]
\begin{center}
\epsfig{figure=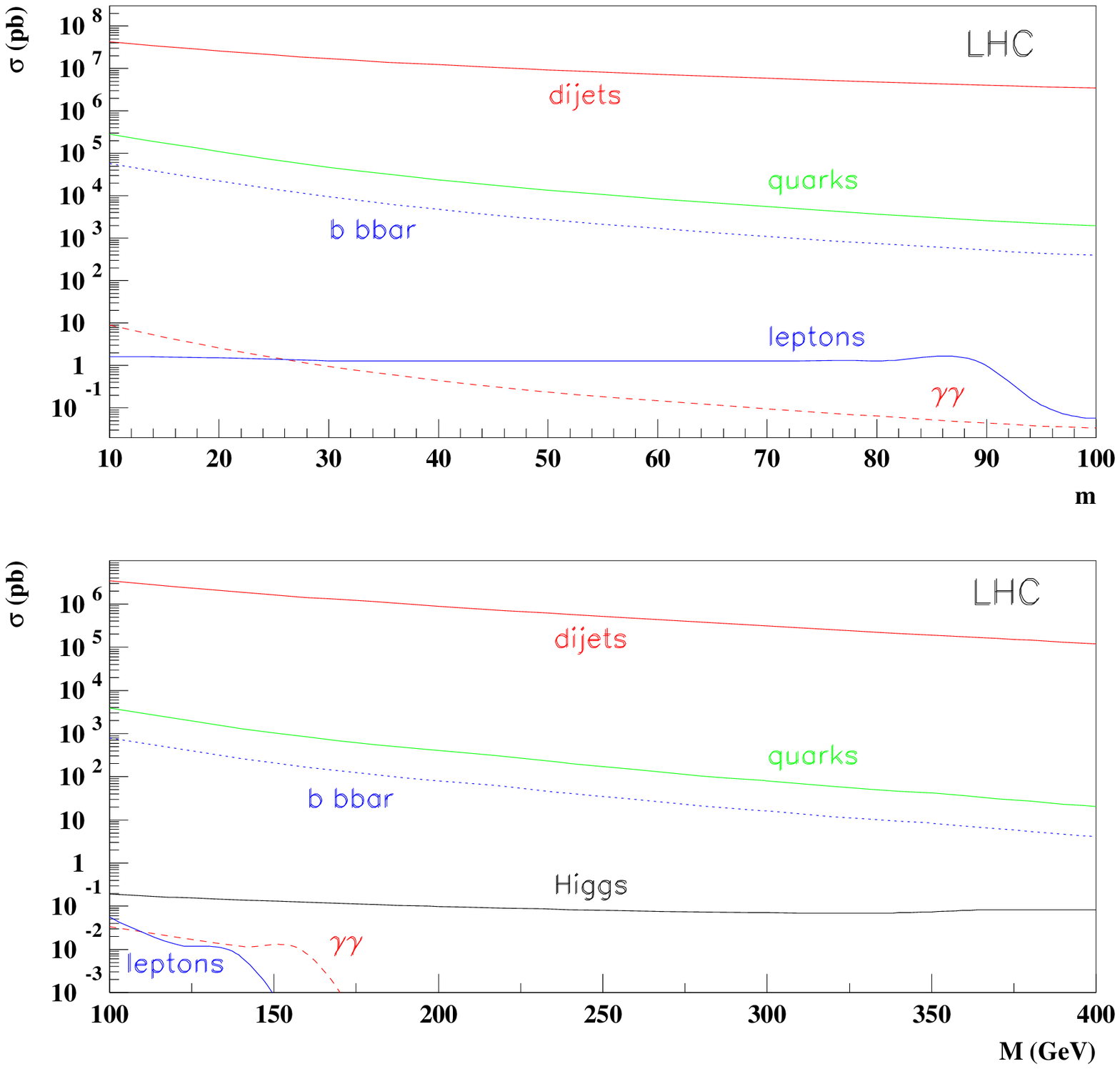,height=6.in}
\end{center}
\caption{Dijet, diquark, $b \bar{b}$, dilepton, and diphoton cross-sections (pb) 
at the LHC. The cross-sections are given for a mass $m$ above the value
on the abscissa, for two different mass ranges. For comparison, we also
display the Higgs boson cross-section for $M_{Higgs}=m$.} 
\label{crosslhcb}
\end{figure}

\begin{figure}[p]

\begin{center}
\begin{tabular}{cc}
\epsfig{figure=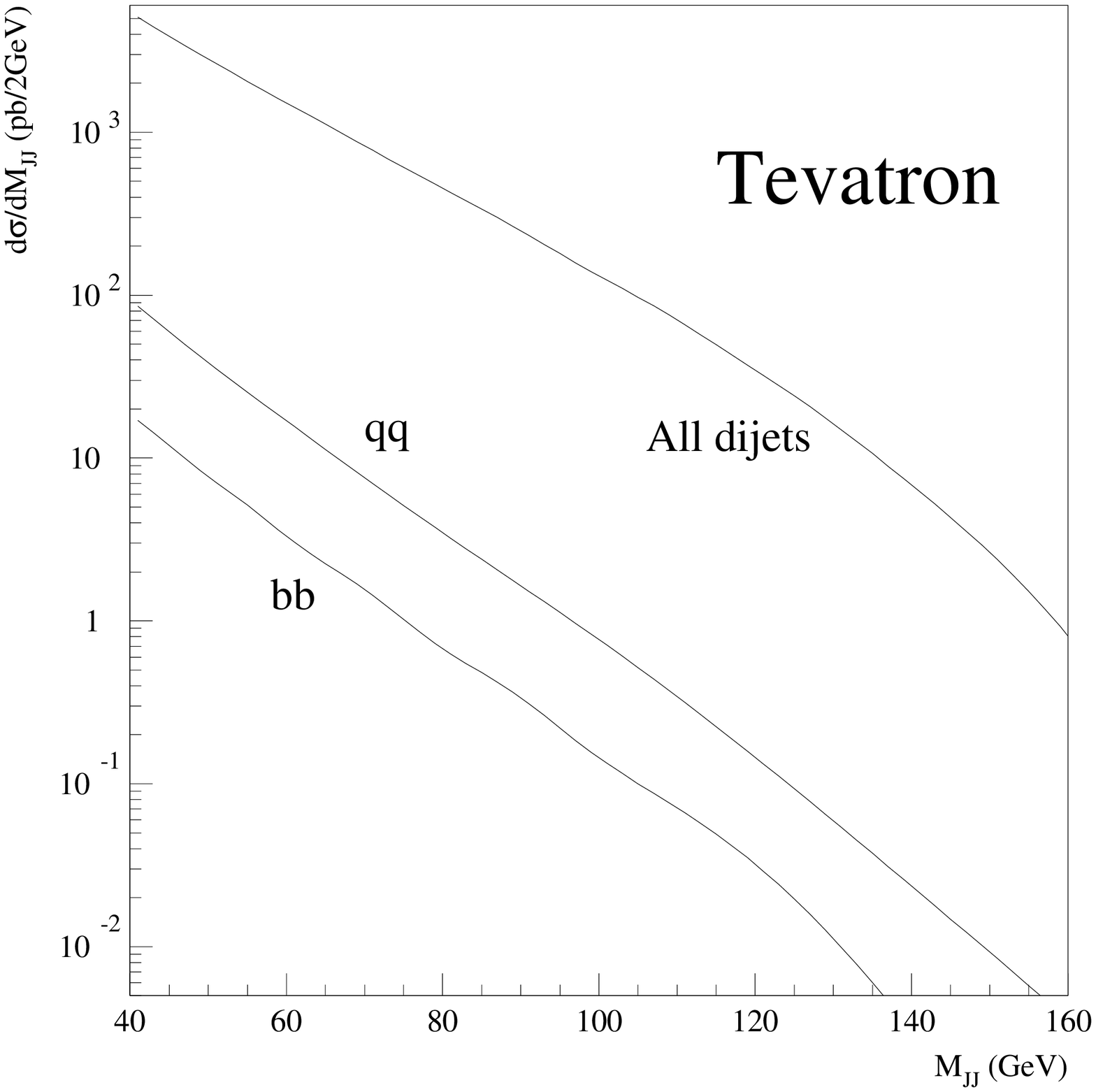,height=2.9in} &
\epsfig{figure=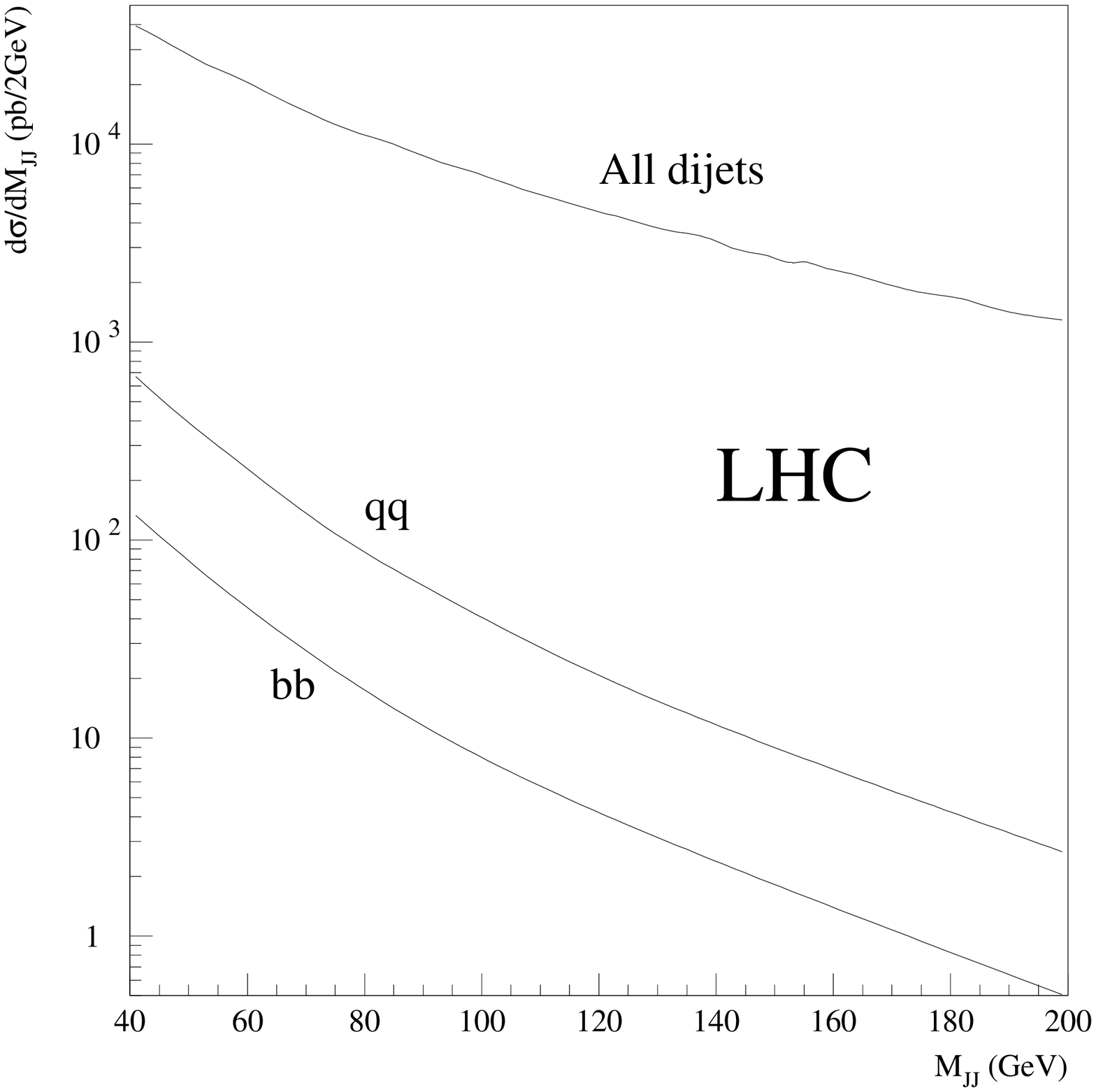,height=2.9in} \\
\end{tabular}
\end{center}
\caption{Differential dijet production cross-section (pb) at the Tevatron and 
the 
LHC. The transverse energy of the central jets satisfies $E_T > 10$ GeV, 
and their rapidity is limited to  $|y|<4$.}
\label{dijetXS}

\begin{center}
\begin{tabular}{cc}
\epsfig{figure=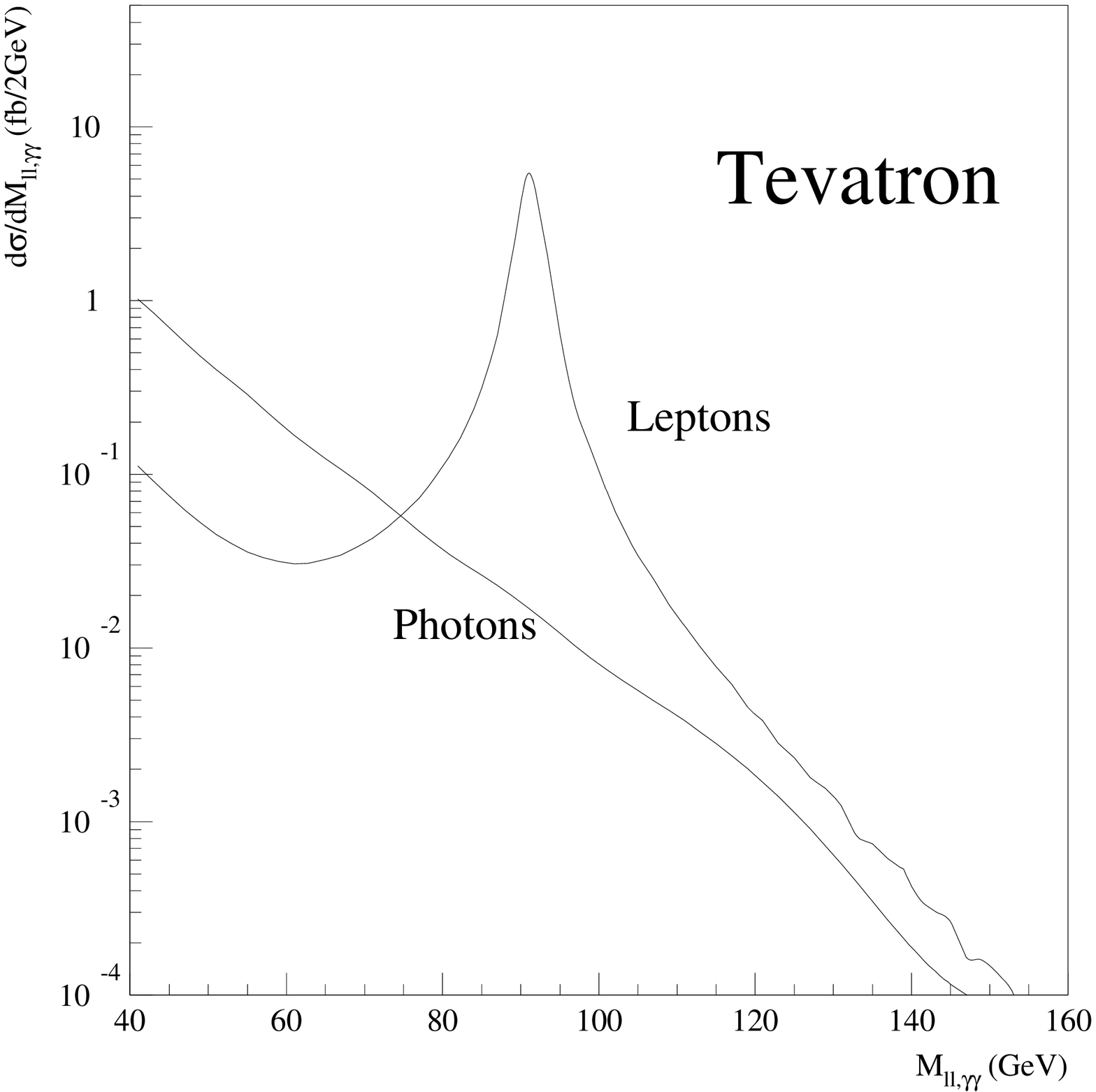,height=2.9in} &
\epsfig{figure=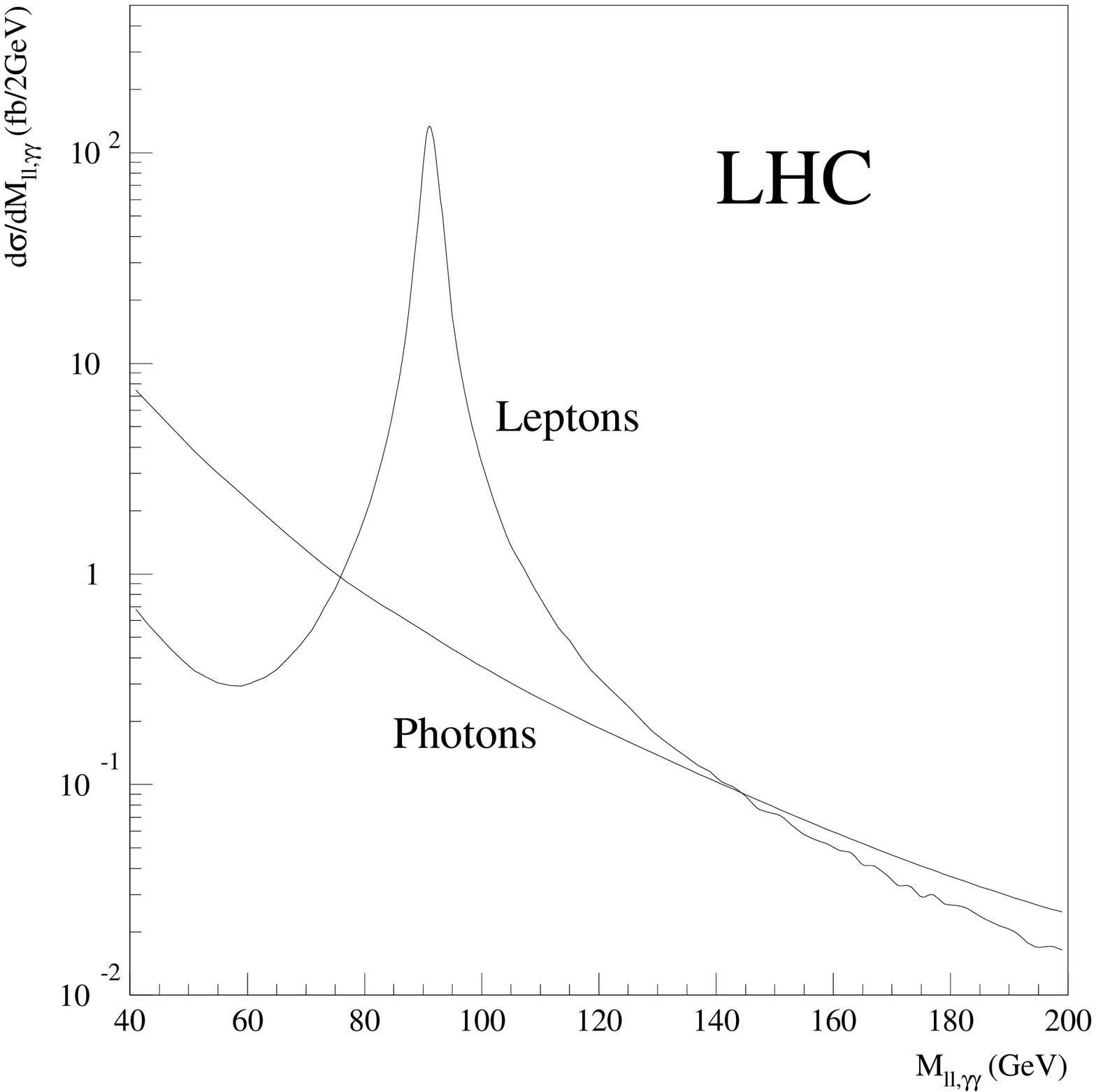,height=2.9in} \\
\end{tabular}
\end{center}
\caption{Differential diphoton and dilepton production cross-sections (fb) at 
the 
Tevatron and the LHC. The 
dilepton cross-section corresponds to a single lepton flavour. The 
transverse energy of 
the central particles satisfies $E_T > 10$ GeV, and their rapidity is 
limited to  $|y|<4$.}
\label{photonXS}
\end{figure}

Even if the diffractive Higgs boson prodiction cross sections are too low at the Tevatron,
important tests of the models concerning dijet, diphoton and dilepton production
can be done, which will allow to make more precise predictions for the LHC.
All models give predictions for these cross sections and we give in Figs 
\ref{crosster}, \ref{crosslhcb}, \ref{dijetXS}, \ref{photonXS} \cite{pesch}
the predictions for the non factorisable inclusive model. All these measurements
will be performed soon by the CDF and D\O\ experiments and will perform good
tests of the model \cite{newpap}.

\begin{figure}[p]
\begin{center}
\epsfig{figure=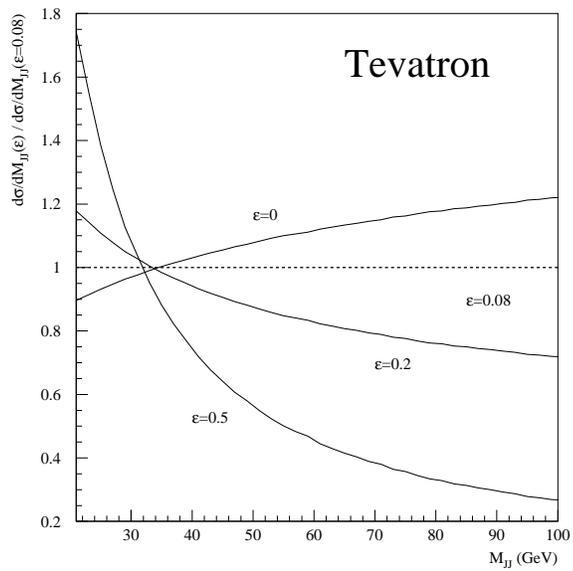,height=3.in}
\end{center}
\caption{Differential dijet cross-sections for different 
values of $\epsilon$, compared to a reference taken at $\epsilon= 0.08$. All 
cross-sections are normalised to the same value. The values $\epsilon= 0.08$, 
0.2 and 0.5 correspond,  
respectively, to the soft, hard (measurement at HERA), and hard BFKL
pomeron intercepts.}
\label{epsilon2}
\end{figure}

\begin{figure}[p]
\begin{center}
\epsfig{figure=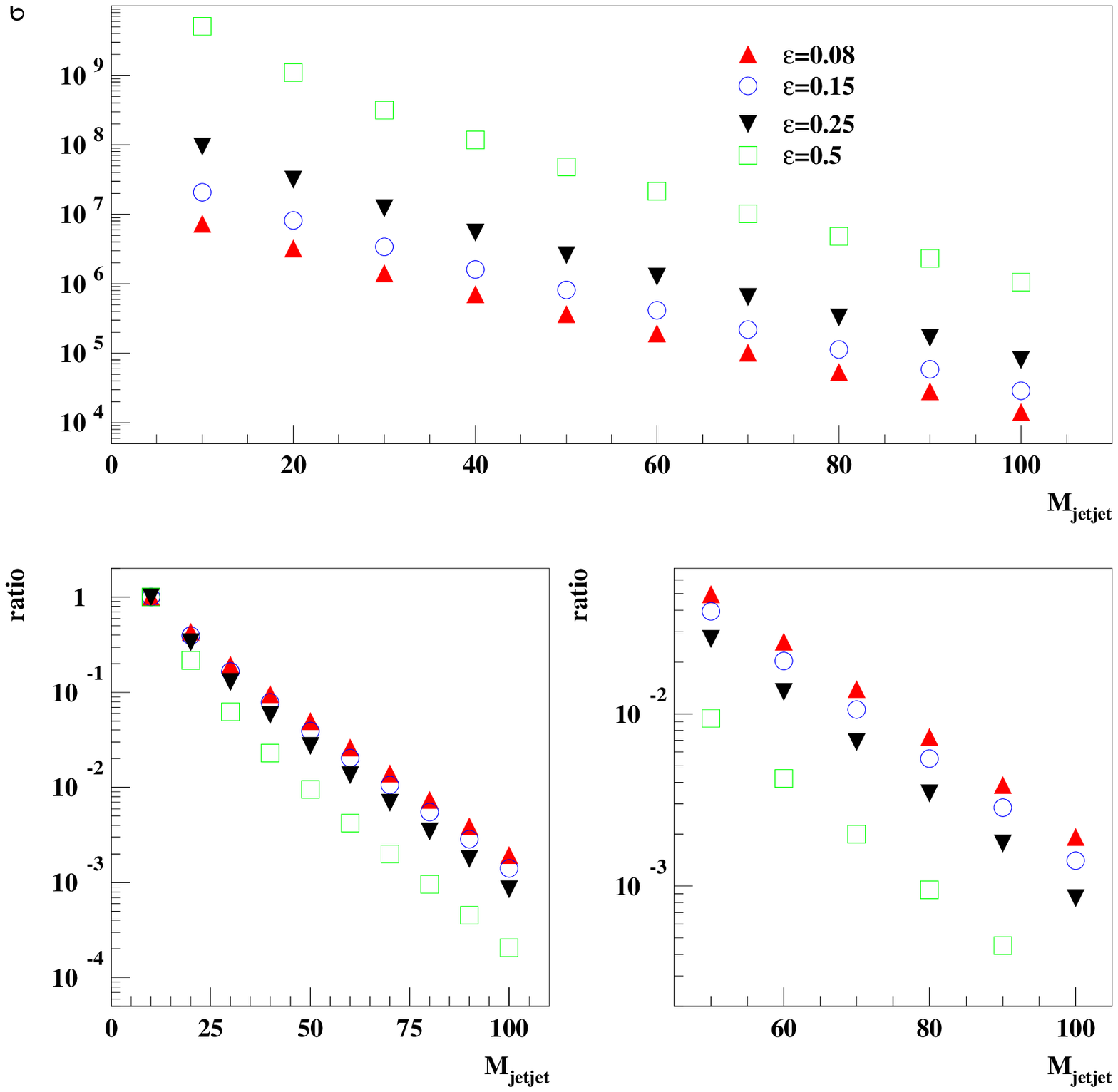,height=5.in}
\end{center}
\caption{Variation of the dijet cross section for different values of
$\epsilon$ as a function of the dijet mass bin. The upper plot gives the 
different dijet cross sections at Tevatron
energies. The lower plots show the same results when we put arbitrarily all
cross section at 1.0 in the first dijet mass bin to show the differences in
slope.}
\label{crossbis}
\end{figure}

It is also important to be able to distinguish between the two inclusive models,
the factorisable and the non factorisable one. As we mentionned before, one of
the main differences between both models is the value of the pomeron intercept
which is the soft one ($\epsilon=0.08$) in the non factorisable model,
and the hard one measured at HERA ($\epsilon=0.2$) in the case of the
factorisable one. Another difference is due to the presence of a soft
gluon between both pomerons in the case of non-factorisable models (see Fig.
\ref{diag2}).
In Fig. \ref{epsilon2} is given the 
differential dijet cross-sections for different 
values of $\epsilon$, compared to a reference taken at $\epsilon= 0.08$. All 
cross-sections are normalised to the same value. We note that the measurement
of $\epsilon$ can be performed using the measurement of the differential dijet mass 
cross section. Fig. \ref{crossbis} also shows that the measurement of the
slope of the dijet mass cross section may lead to a measurement of $\epsilon$
even if this measurement is not easy. Such a measurement will allow to
distinguish between the factorisable and non factorisable models and will
perform a direct measurement of the pomeron intercept at the Tevatron.

\subsection{Roman pot detectors at the Tevatron and the LHC: where to put them?}
\begin{figure}[p]
\begin{center}
\epsfig{figure=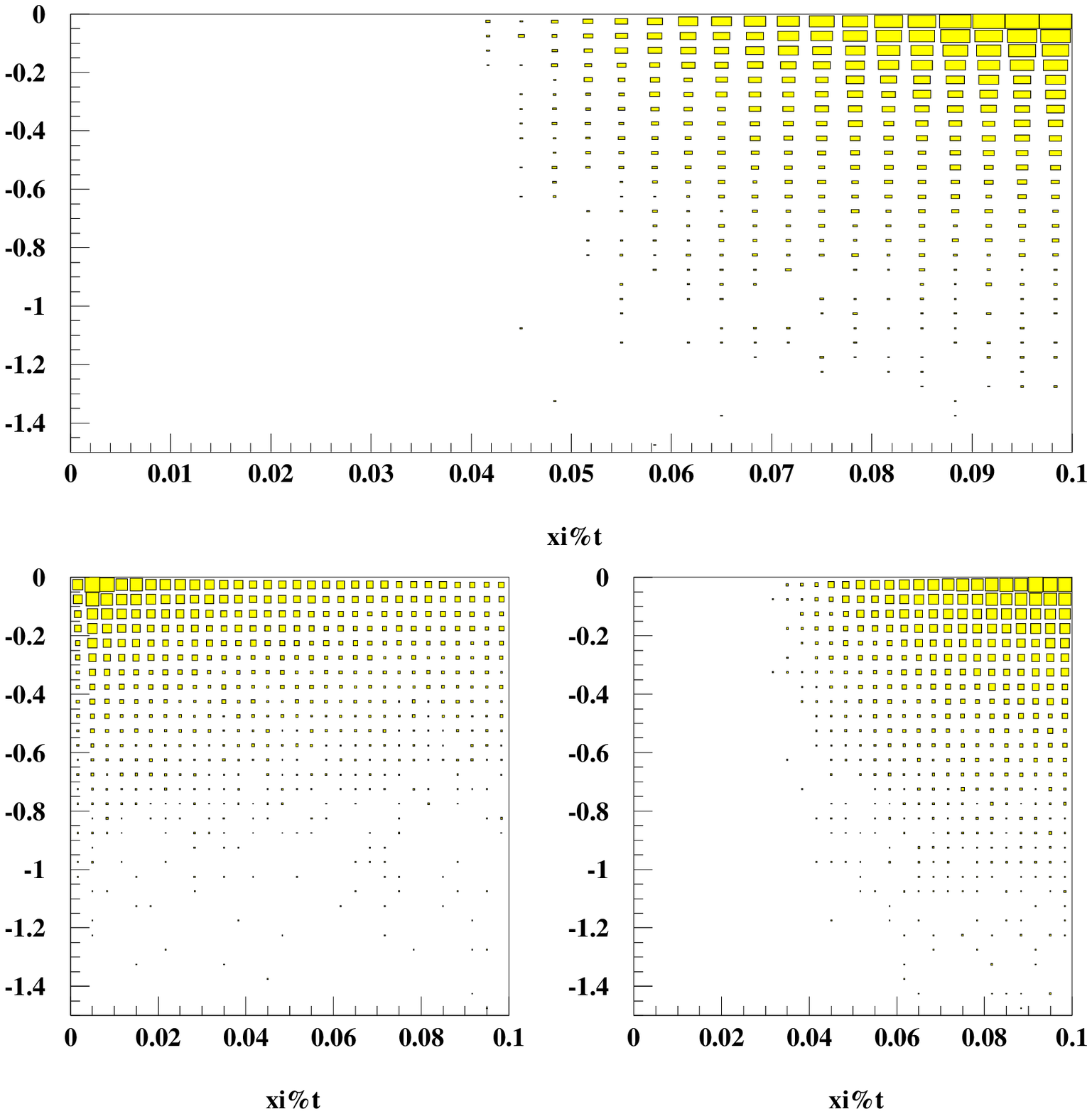,height=4.in}
\end{center}
\caption{$\xi$ vs $t$ distributions for diffractive Higgs boson production.
The size of the squares is proportional to the cross section given by the
non factorisable model. The upper plot give the $\xi$ vs $t$ distribution
at the Tevatron for a Higgs mass of 120 GeV, the lower left plot
for a Higgs mass of 120 GeV at the LHC, and the lower right plot for a Higgs 
mass of 800 GeV at the LHC.}
\label{kinema}
\end{figure}

In this section, we will discuss the optimal positions to detect Higgs bosons
at the Tevatron or the LHC. In Fig. \ref{kinema}, we give the $\xi$ vs $t$ distributions for diffractive Higgs boson production.
The size of the squares is proportional to the cross section given by the
non factorisable model. The upper plot give the $\xi$ vs $t$ distribution
at the Tevatron for a Higgs mass of 120 GeV, the lower left plot
for a Higgs mass of 120 GeV at the LHC, and the lower right plot for a Higgs 
mass of 800 GeV at the LHC (which is basically the same plot as the upper one,
the difference being due only to the difference of beam energies). We notice
that we need a good acceptance at high $t$ ($|t|>0.2$) and high $\xi$
($ \xi > 0.05$) at the Tevatron, and at high $t$ ($|t|>0.2$) and low $\xi$
($ \xi < 0.02$) at the LHC. 

The CMS \cite{cms} and Totem collaborations proposed to install a few sets of roman pot
detectors showing the following acceptances: 
\begin{itemize}
\item (1): 140-180 meter pots (standard TOTEM) $|t|<$ 1.5 
GeV$^2$, $\xi > 0.02$
\item (2): 240 meter pots (addition in warm section), and TOTEM 
$|t|<$ 2 GeV$^2$, $\xi > 0.01$
\item (3): 425 meter pots (cold section) $|t|<$ 2 
GeV$^2$, $ 0.002 <\xi < 0.02$
\item (4): All pots  $|t|<$ 2 
GeV$^2$, $ \xi > 0.002$
\end{itemize}
We notice that we need to have roman pots or microstations \cite{orava} in the
cold region of the LHC if we want to have a good acceptance for diffractive
Higgs production at the LHC. In the following, we will assume that the CMS 
Collaboration has all pots (case number 4) when we discuss the advantage of
reconstructing the Higgs mass using roman pot detectors or microstations
\cite{newpap}.

\section{Exclusive diffractive Higgs boson production}
In this kind of models \cite{martin} a direct perturbative calculation of the
diagram (see \ref{diag1}) is performed using the gluon density in the proton.
This leads to very clean events where the protons are scattered at very small
angle in the beam pipe (and can be tagged in roman pot detectors or
microstations) and the Higgs boson which decays centrally in the main detector,
and nothing else. The problem is that
one needs to suppress totally QCD radiation. The price to pay to get this kind 
of events is an 
exponential Sudakov form factor to suppress QCD radiation which leads to a low 
cross section. One may wonder if these events exist (no soft gluon can be emitted
after interaction), which can be tested at the Tevatron \cite{dinogap}.
The diffractive Higgs boson cross section for this process is given in Table
\ref{martin}. We note that the cross sections are extremely small for the
Tevatron and quite large enough for the LHC provided that the gap survival
probability is not too small.

\begin{figure}[p]
\begin{center}
\epsfig{figure=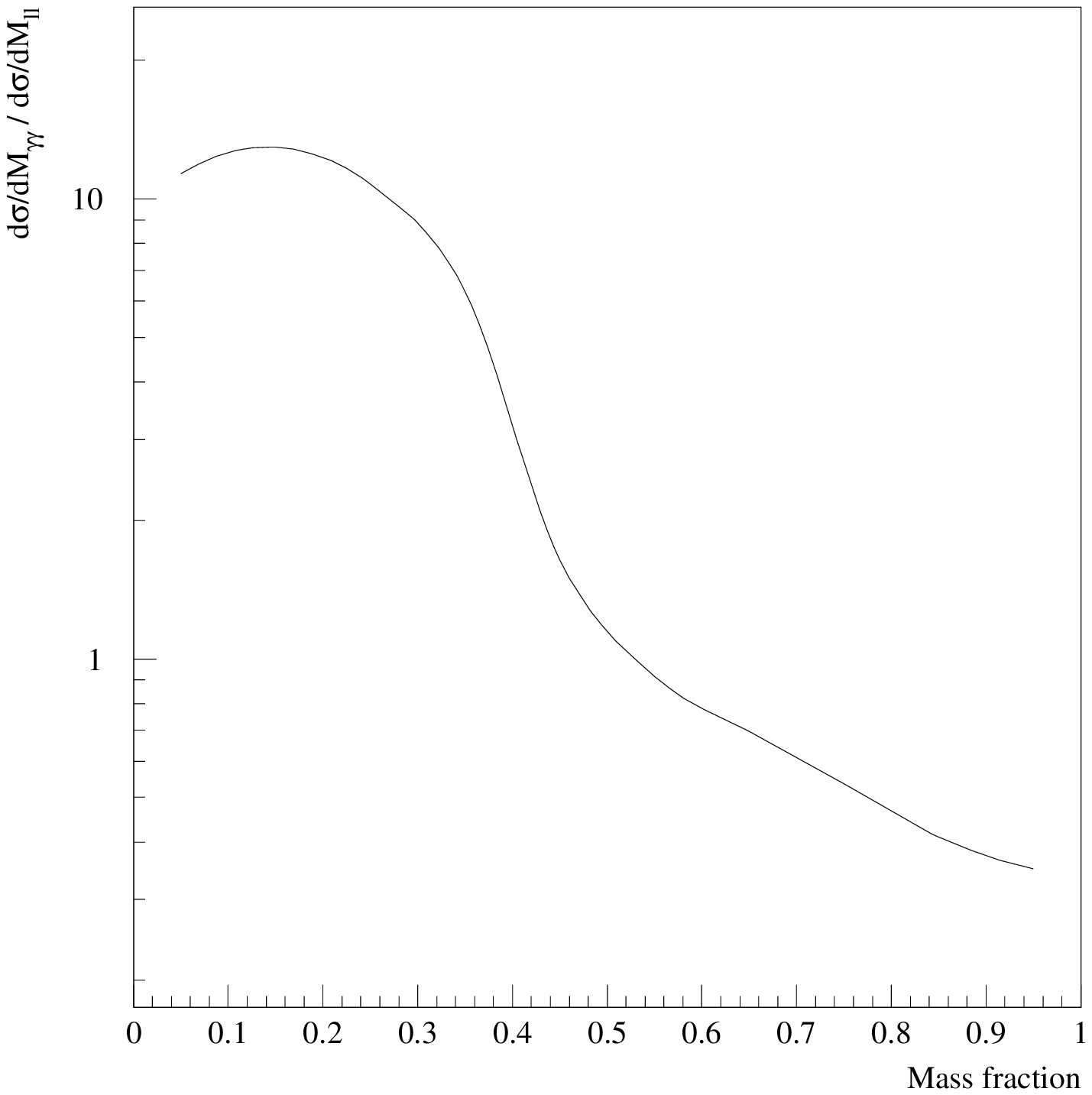,height=3in}
\end{center}
\caption{Diphoton to dilepton cross-section ratio, as a function of the mass 
fraction.}
\label{ggllratio}
\end{figure}

At the Tevatron, it is important to check if the exclusive events exist or not
and to measure their production cross section since they have never been
observed yet. These events would be the ideal ones for the LHC. So far, the CDF
collaboration looked at these events in the dijet channel and put a limit of 3.7
nb on their production cross section \cite{dinogap}. The difficulty is to distinguish these
exclusive events from {\it quasi-exclusive} events produced in inclusive events.
Namely, it is possible to produce events inclusively where most of the energy 
in the pomeron is used to produce dijets, diphotons, dileptons or Higgs
bosons, or in other words, where pomeron remnants show very little energies. These 
`` quasi-exclusive" events will be very similar to the exclusive ones and it is thus  
impoortant to see how one can distinguish between them, and see if exclusive
events exist or not. One way to distinguish between them would be to measure the
ratio of the diffractive diphoton to dilepton cross sections at the Tevatron.
In Fig. \ref{ggllratio}, we give the diphoton to dilepton cross-section ratio, 
as a function of the mass fraction (the diphoton or dilepton mass divided by the
total diffractive mass). In inclusive models, this ratio is determined by the quark
and gluon distributions inside the Pomeron, and the presence of the Z
pole in the dilepton cross-section. In exclusive models, it is only possible to
produce diphoton diffractively but no dilepton. The diphoton cross section obtained
for a mass fraction higher than 0.85-0.9 is of the order of 2 fb for both the
exclusive and inclusive cross sections, 
and the ratio of the diphoton to dilepton
cross section measured at the Tevatron is expected to show an enhancement and a
change of slope as a function of the diphoton/dilepton mass if exclusive events
exist. This will be a very clean test of the existence of exclusive events.
Another possible test will be to measure the $\chi_b$ or $\chi_c$ central
diffractive production cross section.

\begin{table}
\begin{center}
\begin{tabular}{|c||c|c|} \hline
$M_{Higgs}$&Tevatron& LHC  \\
\hline\hline
120 &0.03  &  1.4  \\
140 & 0.02 &  0.9 \\ 
160 & 0.008 &  0.55  \\ \hline
\end{tabular}
\caption{Higgs boson production cross section in fb at the Tevatron and the LHC
for different Higgs boson masses for the exclusive models. 
(see Ref. \cite{martin}
for more detail). (Note that no gap survival probability has been applied
for the LHC).}
\end{center}
\label{martin}
\end{table}

\section{Soft colour interaction models}
In this section, we will discuss a completely different model based
on soft interaction in the final state which happen after the hard interaction
at a longer time scale. 
Two versions of a model have been designed to describe diffractive physics as soft
color rescattering over a hard subprocess \cite{ingelman}, namely SCI (Soft 
Color Interaction)
and GAL (General Area Law). These models do not imply the existence of
colourless objects like the pomeron since diffraction is explained by soft
interactions in the final state. Color exchanges occur in the final state, potentially
stopping color flow between the remnants of the incoming hadrons and
the central system, leading to diffractive event topologies.
They are implemented as a transition
between the hard interaction and hadronization, and therefore fit
naturally in Monte-Carlo programs such as  PYTHIA \cite{pythia}. 

Their prediction concerning diffractive Higgs production are given in Table
\ref{gunnar}. Very low cross sections are obtained at the Tevatron and quite low
cross sections at the LHC (even if they are high enough to be measured at the
LHC because of the high value of the luminosity). The disadvantage of this model is
that it does not describe perfectly final states where a tagged proton is
detected in the final state, especially at HERA.

\begin{table}
\begin{center}
\begin{tabular}{|c||c|c|} \hline
$M_{Higgs}$&Tevatron& LHC  \\
\hline\hline
115 &1.2-2.4 10$^{-4}$  &  0.162-0.189  \\
 \hline
\end{tabular}
\caption{Higgs boson production cross section in fb at the Tevatron and the LHC
for soft colour interaction models. 
(see Ref. \cite{ingelman}
for more detail).}
\end{center}
\label{gunnar}
\end{table}

\section{Higgs mass reconstruction using diffractive events}

\subsection{Higgs mass reconstruction in the case of exclusive events}
\begin{figure}[p]
\begin{center}
\epsfig{figure=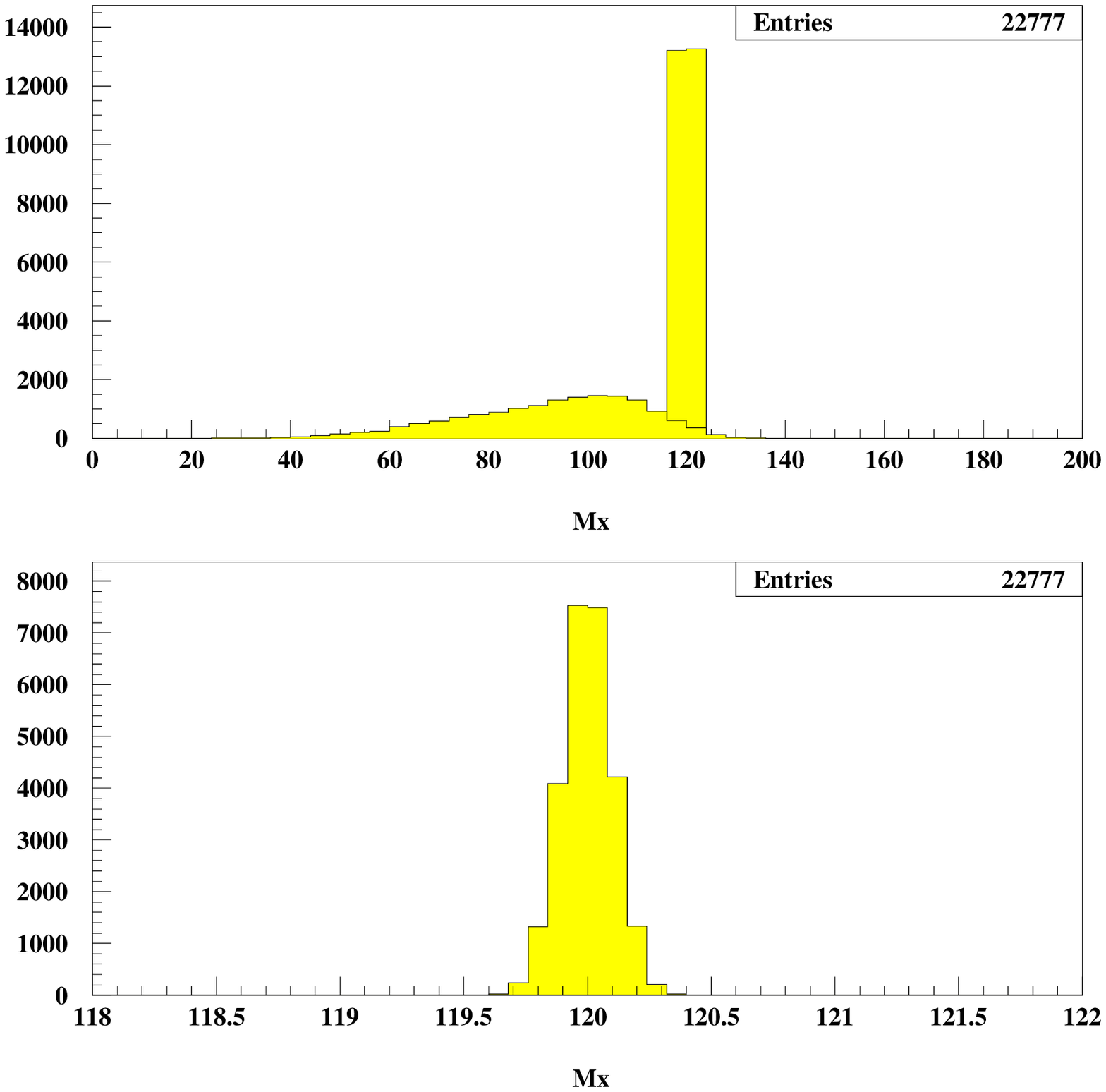,height=4in}
\end{center}
\caption{Higgs mass reconstruction for exclusive events using roman pot
detectors.}
\label{mhiggs}
\end{figure}

Exclusive Higgs events are diffractive events where all the energy (basically all
diffractive mass) is used to produce the Higgs boson. Kinematically, it is
very easy to reconstruct the Higgs mass if one is able to measure the
$\xi$ of both scattered protons detected in roman pot detectors \cite{albrow}:
$M_{Higgs} = \sqrt{\xi_{p1} \xi_{p2} S}$. The mass reconstructed using
roman pot detectors with a resolution in $\xi$ of 0.2\% , and in $t$ of 10\%
$\sqrt t$
is given in Fig. \ref{mhiggs}. We note that these events allow to get a perfect
Higgs mass reconstruction with a resolution better than 1\%. This resolution
allows a very good signal over background separation.
 
The experimental difficulty is to trigger on these events. Since the
microstations or roman pot detectors are located far away from the main
experiments (Atlas or CMS), the trigger informations from these detectors
will only arrive at the second level. At the first level, it is needed to
trigger on energy deposited in the calorimeters requiring for instance the
presence of the Higgs decay products as well as little energy in the forward
region. At the second level, the trigger can require a matching between masses
calculated using the calorimeter or the microstation information. Of course, a
detailled simulation is needed to make more precise statements \cite{newpap}.

\subsection{Higgs mass reconstruction in the case of inclusive events}
\begin{figure}[p]
\begin{center}
\epsfig{figure=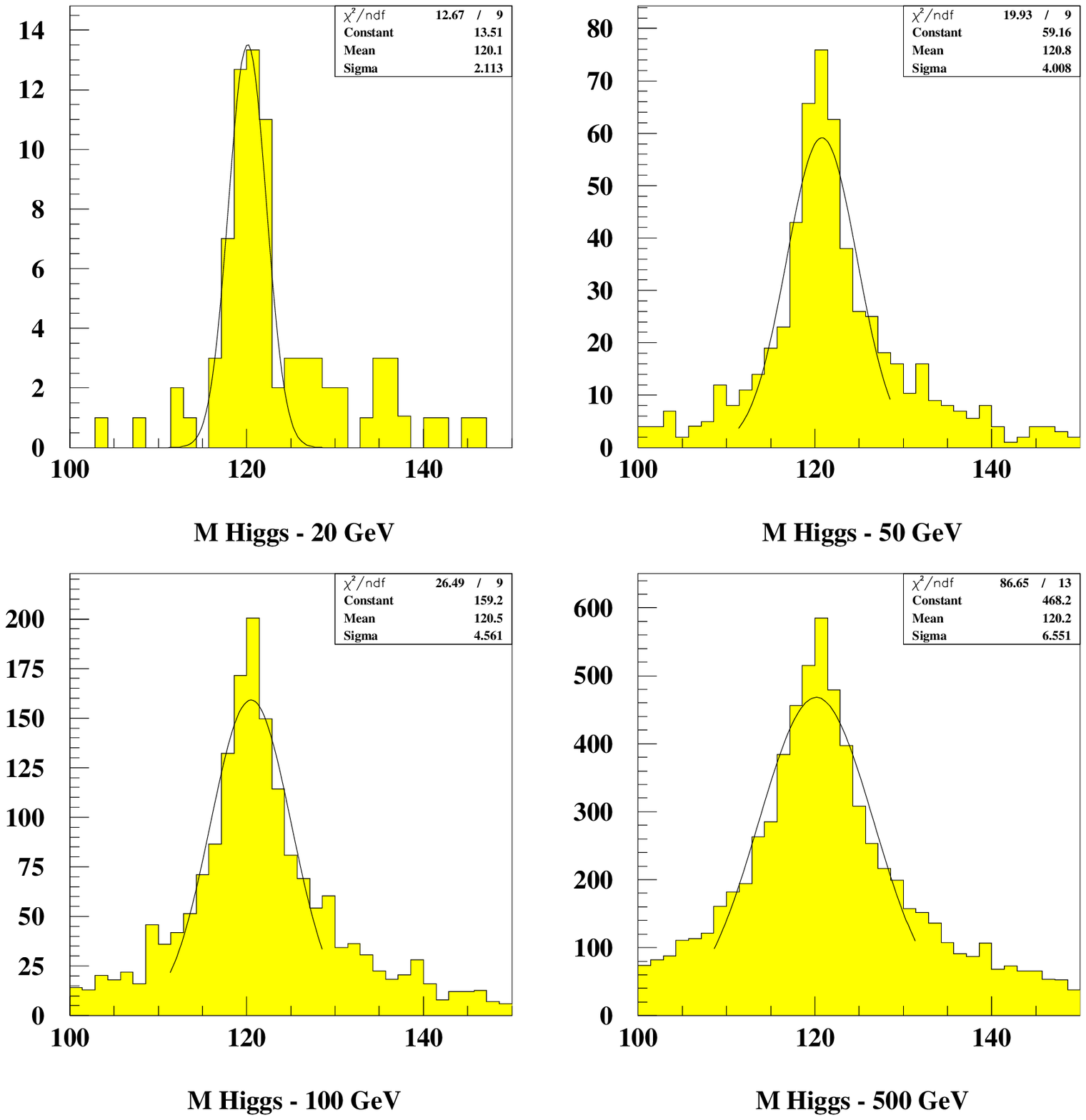,height=4in}
\end{center}
\caption{Higgs mass reconstruction for inclusive events using roman pot
detectors, and central detectors up to a rapidity of 7.5 for a Higgs mass
of 120 GeV at the LHC. }
\label{mhiggsbis}
\end{figure}

In the case of inclusive events, the previous method to reconstruct the mass
of the Higgs boson will not work so nicely since the total energy is used not
only to produce the Higgs boson, but also lost in the pomeron remnants. The
natural idea is to cut on the energy of the pomeron remnants to be able to get
quasi-exclusive events where not much of the available energy is lost in pomeron
remnants. The CMS collaboration \cite{cms} will be able to tag particles up to a
rapidity of 7.5. In Fig. \ref{mhiggsbis}, we give the resolution obtained on the
Higgs mass reconstruction if one is able to cut on the energy of the pomeron
remnant (in these plots, we assume that we are able to tag the remnants up to a
rapidity of 7.5, and the resolution of these detectors to be 100\%$/ \sqrt E$).
The resolution of the Higgs mass is found to be about 2.1, 4.0, 4.6 and 6.6 GeV
if one requires the remnant energies below a rapidity of 7.5 to be respectively
less than 20, 50, 100 and 500 GeV at the LHC. Of course this method will work 
only during the first three years at the LHC when the luminosity will be low
because of pile-up events. Of special interest are the cases when the Higgs
boson decays into $\tau$ because of the low diffractive background.

\section{Conclusion}
In this short review, we have discussed the different models of production of
diffractive Higgs bosons. We can distinguish between three different sets of
models: inclusive, exclusive or soft interaction. All models (even the 
soft colour ones) show low cross
sections at the Tevatron and high enough cross sections at the LHC. A noticeable
enhancement of the production cross section can be obtained in SUSY models
with high values of $\tan \beta$. It is thus very important to test these
models using Tevatron data being taken now, to get more precise cross section
predictions for the LHC.
The Higgs mass reconstruction is greatly
improved by using roman pot detectors or microstations at the LHC, which
enhances the signal to background ratios and allows to use the $b \bar{b}$
and the $\tau \tau$ Higgs
decay channels for exclusive \cite{martin} or quasi-exclusive events 
\cite{newpap}.

To summarise,  we give in Table \ref{final} the cross sections given by the four different
models described in this review both for the Higgs and diphoton
predictions when they are available.

\begin{table}
\begin{center}
\begin{tabular}{|c||c|c|c|c|} \hline
Process &(1)& (2) &  
(3) & (4) \\ 
\hline\hline
$\gamma \gamma$, Tev, $E_T>12 GeV$, $\eta <2$ & 71. & 128. (27.) & - & $\sim$
20.\\
$\gamma \gamma$, Tev, $E_T>12 GeV$, $\eta <1$ & 9. & 8. (2.) & - & - \\
$\gamma \gamma$, LHC, $E_T>50 GeV$, $\eta <2$& 1.5 & - & - & 0.1 \\
$\gamma \gamma$, LHC, $E_T>120 GeV$, $\eta <5$& 19. & - & 0.12 & -\\
\hline\hline
Higgs, 115 GeV, Tev & 1.7 & 0.029-0.092 & 0.2 & 0.00012 \\
Higgs, 115 GeV, LHC & 169. & 379.-486. & 2.8 & 0.19 \\
Higgs, 160 GeV, LHC & 123. & 145. & 1.0 & - \\ \hline
 \hline
\end{tabular}
\caption{Diphoton and Higgs boson cross sections (fb) for the different
models discussed: 1: inclusive non factorisable models,
2: inclusive factorisable models (note that no GSP is used for the LHC),
3: exclusive models,
4: models of soft colour interaction
}
\end{center}
\label{final}
\end{table}

\section*{Acknowledgments}
Most of these results come from a fruitful collaboration with Maarten Boonekamp,
Robi Peschanski and Albert de Roeck.
I also wish to thank Mike Albrow, Valery Khoze, St\'ephane Lavignac, Risto Orava, 
and Laurent Schoeffel for useful
discussions.

\end{document}